\documentclass{article}[12pts]
\usepackage{setspace}
\usepackage[margin=1in]{geometry}
\usepackage{amsmath}
\usepackage{amssymb}
\usepackage{amsfonts} 
\usepackage{bm}
\usepackage{listings}
\usepackage{graphicx}
\usepackage[english]{babel}
\usepackage[utf8x]{inputenc}
\usepackage{tabularx}
\usepackage{adjustbox}
\usepackage{authblk}  
\usepackage[utf8x]{inputenc}
 \usepackage{booktabs} 
\usepackage{subcaption}
\usepackage{xcolor}
\usepackage{hyperref}
\usepackage{float}
\usepackage{array,multirow}
\usepackage{tabularx}
\usepackage{adjustbox}
\usepackage{authblk}
\title{Personalized feature threshold estimation in joint modelling of longitudinal and time-to-event data}
\author[1] {Mirajul Islam}
\author[2]{Michael J. Daniels}
\author[3]{Juned Siddique}
\affil[1]{PhD Candidate, Department of Statistics, University of Florida}
\affil[2]{Professor, Department of Statistics, University of Florida}
\affil[3]{Professor, Department of Preventive Medicine, Northwestern University Feinberg School of Medicine, Chicago}
\date{\today}

\begin{document}

\maketitle
\begin{abstract}
Cardiovascular disease (CVD) cohort studies collect longitudinal data on numerous CVD risk factors including body mass index (BMI), systolic blood pressure (SBP), diastolic blood pressure (DBP), glucose, and total cholesterol. The commonly used threshold values for identifying subjects at high risk are 30 kg/$m^2$ for BMI, 120 mmHg for SBP, 80 mmHg for DBP, 126 mg/dL for glucose, and 230 mg/dL for total cholesterol. When studying the association between features of longitudinal risk factors and time to a CVD event, an important research question is whether these CVD risk factor thresholds should vary based on individual characteristics as well as the type of longitudinal feature being considered.  Using data from the Atherosclerosis Risk in Communities (ARIC) Study, we develop methods to estimate risk factor thresholds in joint models with multiple features for each longitudinal risk factor. These thresholds are allowed to vary by sex, race, and type of feature. Our methods have the potential for personalized CVD prevention strategies as well as better estimates of CVD risk.
\end{abstract}

\section{Introduction}
Cohort studies investigating cardiovascular disease (CVD) frequently collect longitudinal measurements of key biomarkers, including body mass index (BMI), systolic and diastolic blood pressure (SBP and DBP), blood glucose, and total cholesterol (TOTCHL). These biomarkers are well-established risk factors for adverse outcomes such as CVD-related mortality. In both clinical and epidemiological contexts, continuous biomarkers play a critical role in disease diagnosis and prognosis, and are often used to stratify individuals into categories of high, low, or no risk \cite{woo2020determination}. Such stratification supports targeted clinical decision-making, enabling more personalized interventions based on an individual’s risk profile.
However, determining appropriate cutoff points for these biomarkers presents a methodological challenge. Several approaches exist for threshold selection, ranging from data-driven methods to those based on expert consensus or prior literature. Cutoffs may be chosen a priori, informed by biological rationale, clinical expertise, or previously published findings. Nevertheless, reliance on clinical judgment can introduce subjectivity, and cutoffs derived from earlier studies may not generalize well to new populations due to differences in study design, participant characteristics, or outcome definitions \cite{marubini1990long,de1993age}.\\\\
While traditional approaches often apply static thresholds to cross-sectional biomarker values, such methods may overlook the complex, time-varying nature of disease processes. In reality, biomarker trajectories evolve over time, and their association with clinical outcomes, such as cardiovascular events, may depend on the features of these trajectories. In addition, these longitudinal biomarkers are censored by the occurrence of CVD death and are endogenous, as they can be influenced by the event of interest \cite{rizopoulos2012joint}, and are often measured intermittently introducing measurement error. Analyzing these biomarkers separately through linear mixed models for the longitudinal data \cite{laird1982random} and Cox regression models \cite{cox1972regression} for CVD death  can be inefficient and may lead to biased estimates, as the two outcome processes are correlated \cite{ibrahim2010basic}. To improve inference for a time-to-event outcome by taking account of an intermittently and error-prone measured endogenous time-dependent variable, some initial approaches were two-stage models \cite{self1992modeling, tsiatis1995modeling}. However, these approaches ignore the uncertainty and measurement error in the estimated values of the longitudinal process, leading to biased estimates of the association between the covariate and the survival outcome \cite{tsiatis2004joint}.  To address these limitations, joint modeling frameworks that link longitudinal biomarker data with time-to-event outcomes have become a standard alternative over the past two decades \cite{tsiatis2004joint, baart2021joint,henderson2000joint, rizopoulos2014combining, andrinopoulou2016bayesian}. In joint models, the association between time-dependent covariates and event outcomes is often characterized by extracting features from the longitudinal trajectories \cite{rizopoulos2012joint,rizopoulos2014combining}. Commonly used features include the current value, threshold indicator, cumulative area under the trajectory, and the area above a specified threshold \cite{rizopoulos2012joint} etc.\\\\
 Estimating the optimal thresholds of the features of quantitative biomarkers within a joint modeling framework is essential for medical decision-making as it  allows for data-adaptive and potentially personalized cut-off points that account for subject heterogeneity, including differences by sex, race, or biomarker dynamics. This paradigm shift enables a more nuanced understanding of risk and supports individualized clinical decision making based on the longitudinal profile of the patient. Because of complex biomarker distributions due to natural heterogeneity in marker expression and other heterogeneities, methods to estimate a biomarker optimal threshold would be welcome \cite{subtil2014estimating}. When studying the association between features of longitudinal risk factors and time to a CVD event, an interesting research question is whether these CVD risk factor thresholds should vary based on individual characteristics as well as the type of longitudinal feature being estimated.  Using data from the Atherosclerosis Risk in Communities (ARIC) Study, we develop methods to estimate risk factor thresholds in joint models with multiple features for each longitudinal risk factor. These thresholds are allowed to vary by sex, race, and type of feature (current value and area above threshold). Our methods have the potential for personalized CVD prevention strategies as well as better estimates of CVD risk.\\\\
 The ARIC study, a population-based, prospective cohort study consists of over 15,000 individuals' aged 45-64 years at enrollment \cite{atherosclerosis1989atherosclerosis}. Baseline information was collected including medical history, physical activity, medication use, and diet. Participants were re-examined at 6 additional visits from 1990 to 2019.\\\\
 The remainder of the paper is organized as follows. In Section 2, we introduce the framework, including the novel prior for estimating personalized threshold. Sections 3 presents the results from the analysis of the ARIC data set.  Section 4 consists of our recommendations and a discussion.

\section{Estimating Feature Threshold in a joint model}
\subsection{Joint model}
Let $x_{ig}$ be the design vector  of the $g$th risk factor for the fixed-effect regression coefficients ($\bm{\beta}_g$) and  $z_{ig}$ be corresponding design vector for the random effects ($\bm{b}_{ig}$) for $i$th individual. Then, for the $i$th individual, the $g$th longitudinal sub-model is
\begin{eqnarray*}
    y_{ig}(t)&=&\mu_{ig}(t)+\epsilon_{ig}(t); g=1,2,...,G;~i=1,2,\ldots,n,\\
    \text{where}~\mu_{ig}(t)&=&x_{ig}^T(t)\bm{\beta}_g+z_{ig}^T(t)\bm{b}_{ig},
\end{eqnarray*}
	where $\bm{\beta_g}=(\beta_{g0},\beta_{g1},\ldots,\beta_{gp})^T,\epsilon_{ig}\sim N(0,\sigma_g^2).$
To account for the correlation among the $G$ longitudinal
outcomes and also correlation within each longitudinal outcome, it is common to assume a multivariate normal distribution for the corresponding random effects as follows
\begin{equation*}
	\bm{b}_i = (b_{i10},b_{i20},\ldots,b_{iGJ})\sim N(\mathbf{0}, D).
\end{equation*}
 For the survival outcome, let the observed time $T_i=\text{min}~(T_i^*,C_i^*)$ where $T_i^*$ and $C_i^*$ be the event time and censoring time, respectively. Assuming the risk of the event depends on the features of the risk factors and baseline covariates ($w_i$), we write the survival sub-model as
\begin{equation}\label{eq:1}
h_i(t,\bm{\theta_s}) = h_0(t) \exp[w_i^T\bm{\delta}+\sum_{g=1}^{G}\sum_{j=1}^{J}\alpha_{gj}f_{gj}\{\Psi_{ig}(t)\}],
\end{equation}
	where $\Psi_{ig}(t)=\{\mu_{ig}(s),0\leq s\leq t\}$ is the history of the $g$th true unobserved longitudinal process
	up to time point t, $\bm{\delta}$ is the regression coefficients of the baseline covariates, $\alpha_{gj}$  is a set of parameters that link the longitudinal features with survival outcome, and $f_{gj}(.)$ is the $j$th feature of the $g$th risk factor.  Following the approach of Andrinopoulou and Rizopoulos \cite{andrinopoulou2016bayesian}, we specify the baseline hazard function using B-splines,
	$$\log h_0(t) =\gamma_{h_0,0}+\sum_{q=1}^{Q}\gamma_{h_0,q}B_q(t,k),$$
	where $(\gamma_{h_0,0},\gamma_{h_0,1},\ldots, \gamma_{h_0,Q})$ are the vector of spline coefficients and $B_q(t,k)$ denotes the q-th basis function of a B-spline with knots $k_1,\ldots,k_Q$. In equation [\ref{eq:1}], $\bm{\theta_s}=(\bm{\delta},\bm{\alpha},\bm{\gamma}_{\lambda_0})$ is the parameter vector for the survival outcomes. 
Numerous features, $f_{gj}(.)$,  to link the survival process with longitudinal outcomes have been proposed in the literature \cite{rizopoulos2011bayesian, brown2009assessing}. We consider the following two features: current value above threshold, area above threshold.
\begin{eqnarray*}
     f_{g1}\{\Psi_{ig}(t)\}&=&\mu_{ig}(t)I(\mu_{ig}(t)>\gamma_g^{v})\hspace{2.1cm}[\text{threshold}]\\
		f_{g2}\{\Psi_{ig}(t)\}&=&\int_{t_0}^t\mu_{ig}(s)I(\mu_{ig}(s)>\gamma_g^{a})ds\hspace{0.1cm}[\text{area above threshold}]
\end{eqnarray*}
So, the  hazard sub-model specified in [\ref{eq:1}] in our case is
\begin{eqnarray}\label{E:hazard}
    h_{i}(t \vert w_i, f_{i}(t)) &=& h_{0}(t)\exp\big[\delta^{T}w_{i} + \sum_{g=1}^G\alpha_{g1} \mu_{ig}(t)I(\mu_{ig}(t) > \gamma_g(l))+\\
    &&\hspace{3cm}\sum_{g=1}^G\alpha_{g2}\int_{s=0}^{t} \mu_{ig}(s) I(\mu_{ig}(s) > \gamma_g(l))ds \big],\nonumber
\end{eqnarray}
where $\gamma_g(l)$ is the $g$th  risk factor threshold for the  $\ell$th combination of baseline covariates and feature type.  We want to model $\gamma_g$ as a function of time-invariant covariates like race (White=1, Black=0), and sex (Male=1, female=0). Since race, and sex have 4 combinations, we will have 8 thresholds for each longitudinal outcome with feature type (FT) (current value=1, area=-1) i.e.,
\begin{equation*}
  \gamma_{g}(l)=
      \gamma_{g}(\text {FT,SEX,RACE}). 
\end{equation*} 
That is, we are also allowing the thresholds to vary by feature type. 
It is important to note that the features within a risk factor are often correlated.  So, we consider the prior for the association parameters carefully.

\subsection{Prior formulation for threshold estimation}
The identification of a threshold for a feature derived from a longitudinal risk factor within a joint modeling framework requires that both the feature and the underlying risk factor contribute meaningfully to the hazard component of the model. In other words, a threshold is only estimable when the inclusion of both components significantly improves the fit of the survival submodel. Furthermore, even if a particular feature is associated with an increased hazard of a cardiovascular disease (CVD) event, this does not necessarily imply the presence of a distinct threshold effect. A feature may influence the hazard in a continuous or linear fashion without exhibiting a change-point or abrupt increase in risk. To appropriately address these modeling subtleties, we construct our prior specification to account for the possibility that (i) no threshold exists, (ii) the feature is not relevant to the hazard, or (iii) the threshold is dependent on the joint presence of the risk factor and its derived feature in the model.
\\\\
Recall $\gamma_g^j(l)$ is the jth feature threshold for the gth risk factor  at the covariate level ($\ell$) defined by the combinations of sex and race. Let $\sigma_g$ be the standard deviation of the $g$th risk factor. First, we consider the importance of the feature itself, 
\begin{eqnarray*}
\gamma_g^j(l)=\begin{cases}
\text{NA}~~~~~~~~\text{if}~~\alpha_{gj}=0\\
\gamma_g^j(l)^*~ ~~~~\text{if}~~ \alpha_{gj}\neq 0\\
\end{cases}
	\end{eqnarray*}
    where NA denotes no threshold can be identified since the feature is unrelated to the hazard.
    If the feature is related to the hazard, it may or may not have a threshold. So, we specify a spike and slab type prior conditional on $\alpha_{gj}\neq 0$,  
    \begin{equation}\label{eq:2}
       \gamma_g^j(l)^*|\alpha_{gj}\neq 0\sim \pi_{gj} I(\gamma_g^j(l)^*=\text{NA})+(1-\pi_{gj})p(\gamma_g^j(l)^*). 
    \end{equation}
If a threshold exists, it could depend on the baseline covariates and vary by feature type. As such, we specify a prior for $\gamma_g^j(l)^*$ given threshold exists in [\ref{eq:2}] which shrinks the personalized threshold to an additive function  of the baseline covariates considered and feature type (for stability),
\begin{equation}\label{eq:3}
    p(\gamma_g^j(l)^*)=N(m_{0g}+m_{g}^F FT+m_{g}^S Sex+m_{g}^R Race,1/\tau_g^2).
\end{equation}
 We specify the priors on regression coefficients in [\ref{eq:3}]as follows,
 \begin{equation}\label{eq:coeff}
    m_g^{F}\sim N(0,C\sigma_g);~~m_g^{S}\sim N(0,C\sigma_g);~~m_g^{R}\sim N(0,C\sigma_g), 
 \end{equation}
where multiplier $C$ is a constant (usually less than 1) as we do not expect large deviations based on the risk factor sd. The recommended thresholds ($gv$) in the literature (\cite{franklin2009single,grotto2006prevalence}) for the risk factors we consider are BMI, 30; SBP, 120; DBP, 80; glucose, 126; and total cholesterol, 230. For the $g$th risk factor, we set the population average thresholds to this recommended value,
$${gv}_g=\sum_{Sex,Race}\big(m_{0g}+m_{g}^S Sex+m_{g}^R Race\big)p(Sex,Race),$$
where $p(\cdot)$ denotes the empirical distribution. The effect of feature type is canceled out in the specification of ${gv}_g$ because it takes the value 1 for current value threshold feature, -1 for area above threshold feature.
 We can rewrite the above equation in terms of intercept, $m_{0g}$, in \ref{eq:3}
 \begin{equation}\label{eq:4}
     m_{0g}={gv}_{g}-w_1m_{g}^S-w_2m_{g}^R,
 \end{equation}
where $w_i$'s are the weights. To calculate $w_1$ and $w_2$, let $X_1$ and $X_2$ be the binary variables for sex and race. The parameters of the joint distribution $p(X_1,X_2)$ is defined as,
$$p_{lm}=P(X_1=l,X_2=m);~~l,m=0,1.$$
Let
\begin{equation}\label{eq:5}
m_{g0}^*=\sum_{X_1,X_2}\big(m_{0g}+m_{g}^S X_1+m_{g}^R X_2\big)p(X_1,X_2).
\end{equation}

The \ref{eq:5} can now be simplified to
\begin{eqnarray*}
    m_{g0}^*&=&m_{0g}+(p_{11}+p_{10})m_{g}^S+(p_{11}+p_{01})m_{g}^R\\
    m_{0g}&=&m_{g0}^*-w_1m_{g}^S-w_2m_{g}^R
\end{eqnarray*}
So we do not need to specify a prior for $m_{0g}$. Finally, we specify
$$\tau_g^2\sim Gamma(1,1).$$
\subsection{Prior for association parameters}
We want to select important risk factors of CVD as well as important features within a risk factor. For this bi-level selection problem, Xu and Ghosh \cite{xu2015bayesian} proposed Bayesian sparse group selection with spike and slab priors (BSGS-SS). However, this method does not take into account the correlation within each group (risk factor). Islam et. al. \cite{islam2024bayesian} modified the BSGS-SS prior with a Dirichlet hyperprior (DP), namely BSGS-D prior, to select features within each risk factor to better reflect scientific understanding. We use BSGS-D priors for $\alpha_g$. We review the specification here. The coefficient vectors of the feature are parameterized as,
	$$\bm{\alpha}_g = \bm{V}_g^{\frac{1}{2}}\bm{d_g}, \text{where}~ \bm{V}_g^{\frac{1}{2}}=diag\{\tau_{g1},\tau_{g2},\ldots,\tau_{gJ}\}, \tau_{gj}\geq 0.$$
 To select variables at the risk factor level, we specified the following multivariate spike and slab prior for each $\bm{d_g}$,
	$$\bm{d}_g\overset{ind}{\sim}(1-\pi_{g}) N_{J}(\bm{0}, \bm{I}_{J}) + \pi_{g}\delta_0 (\bm{d_g}),~~~ g = 1,2,\ldots,G,$$
	where  $\delta_0 (\bm{d_g})$ denotes a point mass at $\mathbf{0}\in\mathbb{R}^{J}$. The diagonal elements of $\bm{V}_g^{\frac{1}{2}}$
	control the magnitude of elements of $\bm{\alpha}_g$. Note that when $\tau_{gj} = 0, \alpha_{gj}$ is dropped out of the model even when $d_{gj}\neq 0$.
In order to choose features within each relevant risk factor, the following spike and slab prior for each $\tau_{gj}$ is introduced
	$$\tau_{gj}\overset{ind}{\sim}(1-\pi_{gj})N^+(0,s^2)+\pi_{gj}\delta_0(\tau_{gj}),$$
	where $N^+(0,s^2)$ denotes a normal distribution with mean 0 and variance $s^2$ truncated below at 0 and an inverse gamma prior for $s^2$,
 \begin{equation*}
      \frac{1}{s^2}\sim~\text{Gamma}~(1,t)
 \end{equation*}
where $t$ is the scale parameter. Conjugate beta hyper-priors were assumed for $\pi_g$,
$$\pi_g\sim \text{Beta}(a, b).$$
The specification takes into account that it is unlikely to select more than one or two features for a given risk factor by specifying a Dirichlet prior to induce correlation among the $\pi_{gj}$ for each $g$ that will implicitly penalize selecting too many features. For the $g$th risk factor with J features,  define the following probabilities,
	\begin{eqnarray*}
q_{gc}=\begin{cases}
\text{the probability of selecting one feature for}~c=1,\ldots,J\\
\text{the probability of selecting two features for}~c=J+1,\ldots,\frac{J(J+1)}{2}\\
\vdots\\
\text{the probability of selecting all J features for}~c=C
\end{cases}
	\end{eqnarray*}
where $C=2^J-1$, then it specifies Dirichlet prior as,
$$(q_{g1},q_{g2},\ldots,q_{gC})|a_{g1},a_{g2},\ldots,a_{gJ}\sim~Dirichlet(\bm{a_{g}}),$$
where
$$\bm{a_g}=(\underset{\{J~ times\}}{ a_{g1},\ldots,a_{g1}},\underset{\{\frac{J(J-1)}{2}~ times\}}{a_{g2},\ldots,a_{g2}},\ldots,a_{gJ});~a_{g1}>a_{g2}\ldots> a_{gJ}>0.$$

\subsection{Posterior Sampling}
The posterior distribution of the fixed effects, random effects, covariance matrix, survival model parameters, and threshold parameters is given by, 
$$p(\bm{\theta},\bm{b}| y_{i}, T_i,\Delta_i)\propto\prod_{g=1}^G\prod_{\ell=1}^{n_i}p(y_{i\ell g}|b_{ig},\theta_{y_g})p(T_i,\Delta_i|\Psi_{ig}(T_i),\theta_s)p(b_{ig}|\theta_{y_{g}})p (\theta_{y_g})p(\theta_s),$$
where $\bm{\theta}=(\bm{\theta}_s^T,\bm{\theta}_{y_g}^T,\bm{D})~\text{and}~\bm{\theta}_{y_g}=(\bm{\beta}_g^T,\sigma_g),\bm{\theta}_s=(\bm{\delta}^T,\bm{\alpha_1}^T,\bm{\alpha_2}^T,\ldots,\bm{\alpha_G}^T,\gamma_g^j)$.
 We sample from the posterior distribution using NIMBLE \cite{nimble-article:2017} in R studio \cite{team2018rstudio}.
\section{Results}
We used our approach to find the important features and personalized thresholds of CVD risk factors using data from the ARIC study. Posterior estimates, standard deviation (SD), and 95\% credible interval (CI) for the survival parameters in [\ref{E:hazard}] are presented in Table \ref{tab:surv}. The percentages of selection for the association parameters are also shown in the table. We found that both the value and the area features of BMI are important. The area feature of SBP and the value feature of glucose are also important as these features are selected 100\% of the time. Each of these features is positively associated with the risk of CVD death, implying they increase the hazard. The results also show that race, sex, and smoking status have significant effects on the hazard of CVD death. White individuals have an estimated approximately 43\% lower hazard of CVD compared to Black individuals. The hazards of CVD are higher among males compared to females and smokers compared to non-smokers.\\
\begin{table}[H]
	\centering
	\caption{Survival parameter estimates with 95\% credible intervals and percentage of selection}
    \resizebox{\textwidth}{!}{%
	\begin{tabular}{c|c|c|c|c|c}
    \hline
    Variables&Estimate&SD&2.5\% CI&97.5\% CI&\% of selection\\
		\hline
        Value-BMI&-0.026&0.005&-0.035&-0.017&100\\
        Area-BMI*&0.008&0.001&0.006&0.009&100\\
        Value-SBP&0.000&0.001&-0.001&0.004&14\\
        Area-SBP*&0.002&0.000&0.002&0.003&100\\
        Value-Glucose&0.008&0.001&0.006&0.009&100\\
        Area-Glucose*&0.000&0.000&0.000&0.001&47\\
        Race&-0.558&0.137&-0.820&-0.287&-\\
        Sex&0.658&0.116&0.441&0.889&-\\
        Smoking&0.841&0.046&0.750&0.932&-\\
        \hline
        \multicolumn{4}{l}{\small * indicates results in that row were multiplied by 10 except last column.} \\
	\end{tabular}
    }
	\label{tab:surv}
\end{table}
Table \ref{tab:thld} shows the posterior threshold estimates with the corresponding  95\% CIs by race / sex strata for each type of feature. We define significant difference in threshold between two groups based on the CIs of the threshold difference. For the value feature threshold of BMI, black female individuals have significantly higher threshold than white female individuals. Similarly, the threshold of the area feature for BMI is higher among black males than white males. The lowest threshold for the area feature of SBP is obtained among white female individuals. This group has significantly lower threshold than other groups. White individuals have a higher glucose value feature threshold than black individuals though this difference is not significant. The threshold estimates and 95\% CIs are stable for C=0.2,0.3 and 0.4 in \ref{tab:thld}. 
\begin{table}[H]
	\centering
	\caption{Threshold estimates with 95\% credible intervals for different values of C for the prior in \ref{E:hazard}}
  \resizebox{\textwidth}{!}{%
	\begin{tabular}{c|c|c|c|c|c|c|c|c|c|c|c}
		\hline
		&&&\multicolumn{3}{c|}{C=0.2}& \multicolumn{3}{c|}{C=0.3}& \multicolumn{3}{c}{C=0.4}\\
		\hline
		Feature&SEX&RACE&Estimate&2.5\%&97.5\% &Estimate&2.5\% &97.5\%&Estimate&2.5\% &97.5\%\\
        \hline
        \multirow{4}{*}{Value-BMI}&\multirow{2}{*}{Male}&White&22&21&26&22&20&28&22&20&26\\
        &&Black&22&15&25&22&18&25&22&17&25\\
        \cline{2-12}
         &\multirow{2}{*}{Female}&White&20&18&23&21&18&23&21&18&23\\
          &&Black&26&24&29&26&20&32&26&23&29\\
		\hline
        \multirow{4}{*}{Area-BMI}&\multirow{2}{*}{Male}&White&25&17&31&31&23&33&27&23&32\\
        &&Black&35&32&38&35&33&38&35&32&38\\
        \cline{2-12}
         &\multirow{2}{*}{Female}&White&32&30&34&32&29&34&32&30&34\\
          &&Black&31&29&36&34&29&38&33&29&37\\
		\hline
        \multirow{4}{*}{Area-SBP}&\multirow{2}{*}{Male}&White&140&134&146&138&132&144&138&132&145\\
        &&Black&144&130&155&145&135&155&148&137&156\\
          \cline{2-12}
         &\multirow{2}{*}{Female}&White&119&113&124&121&115&128&122&115&128\\
          &&Black&145&139&151&144&134&150&143&136&150\\
		\hline
        \multirow{4}{*}{Value-Glucose}&\multirow{2}{*}{Male}&White&119&113&124&116&109&122&118&113&124\\
        &&Black&112&104&120&110&104&121&112&104&120\\
          \cline{2-12}
         &\multirow{2}{*}{Female}&White&124&117&130&123&116&128&124&118&130\\
          &&Black&118&111&126&118&109&127&119&111&128\\
		\hline
	\end{tabular}
 }
	\label{tab:thld}
\end{table}
\section{Discussion}
In this manuscript, we proposed a novel prior within a joint modeling approach to identify important longitudinal features of CVD risk factors and to estimate their associated thresholds. Our prior allows us to determine whether these CVD risk factor thresholds vary based on individual characteristics as well as the type of longitudinal feature being estimated. We formulated our prior taking into account the identifiability issue of the thresholds. Since the features within a risk factor are often correlated, we used BSGS-D prior that can efficiently deselects the unimportant feature within a risk factor \cite{islam2024bayesian}. We considered two features (current value and area) for each longitudinal risk factor.\\\\
Among the longitudinal biomarkers evaluated, both the current value and area of BMI were found to be strongly positively associated with increased CVD hazard. In addition, the area feature of SBP and the value feature of glucose were consistently selected as important predictors. The consistent selection and positive associations of these features with CVD mortality underscore the importance of considering both instantaneous values and cumulative exposure when modeling risk factor effects over time.\\\\
In terms of baseline covariates, we observed significant differences in CVD hazard by race, sex, and smoking status. Specifically, white individuals had significantly lower hazard of CVD death compared to black individuals. Similarly, males and smokers exhibited significantly higher risks compared to females and non-smokers, respectively—findings that align with established epidemiological evidence (\cite{us2014health,maas2010gender}).\\\\
A key innovation of our approach is the estimation of individualized thresholds for longitudinal features. The threshold estimates revealed important subgroup differences. For example, black female individuals had significantly higher thresholds for the value feature of BMI compared to their white counterparts, suggesting potential variation in risk stratification needs by race and sex. Similarly, black males had higher area-based BMI thresholds than white males. For SBP, the lowest area thresholds were observed among white females, indicating potential heightened sensitivity to SBP burden in this subgroup.\\\\
These findings demonstrate the utility of threshold-based feature modeling in capturing potentially individualized risk relationships between biomarkers and CVD outcomes. By allowing thresholds to vary by demographic covariates, our model supports more tailored risk stratification strategies, which may enhance the precision of prevention and intervention efforts in cardiovascular epidemiology.\\\\
Future work should explore the clinical utility of these thresholds in risk prediction and intervention planning, including their potential integration into decision-support systems. In addition, replication of our findings in external cohorts and evaluation of additional biomarkers will help validate the generalizability and robustness of this approach.\\\\
\hspace{-0.5cm}\textbf{Acknowledgments}\\
All authors were partially supported by NIH R01 HL 158963. ARIC data are available through the NHLBI BIOLINCC data repository.
\bibliography{literature}
\end{document}